# Snowmass 2021 Process

## *White Paper on **L**eading-**E**dge technology **A**nd **F**easibility-directed (**LEAF**) Program aimed at readiness demonstration for Energy Frontier Circular Colliders (pp, $\mu\mu$) by the next decade*

March 14th 2022

G. Ambrosio, G. Apollinari*, M. Baldini, R. Carcagno, C. Boffo, B. Claypool, S. Feher, S. Hays, D. Hoang, V. Kashikhin, V.V. Kashikhin, S. Krave, M. Kufer, J. Lee, V. Lombardo, V. Marinozzi, F. Nobrega, X. Peng, H. Piekarz, V. Shiltsev, S. Stoynev, T. Strauss, N. Tran, G. Velev, X. Xu, A. Zlobin
*Fermi National Accelerator Laboratory, Batavia, IL 60510*
*Corresponding Author: apollina@fnal.gov

K. Amm, M. Anerella, A. Ben Yahia , R. Gupta, P. Joshi, B. Parker, J. Schmalzle,
*Brookhaven National Laboratory, Upton, NY 11973*

P. Ferracin, I. Pong, S. Prestemon, X. Wang, G. Sabbi, T. Shen
*Lawrence Berkeley National Laboratory, Berkeley, CA 94720*

L. Cooley
*Florida State University and NHFML, Tallahassee, FL 32310*

J. Rochester, M.D. Sumption
*The Ohio State University, Columbus, OH 43210*

**Table of Contents**

# 1. EXECUTIVE SUMMARY

Magnets at the limit of - or beyond - existing technical capabilities will be necessary in future HEP facilities being considered by the community at this time, such as muon colliders or next generation high energy hadron colliders.

Historically, the development and demonstration of maturity of advanced magnet technology for application in the present upgrades to the LHC (called the High Luminosity LHC Upgrade, HL-LHC) was made possible by a ~15 years-long US national program of *directed R&D* (called LHC Accelerator Research Program, LARP) working in combination with *generic* and *complementary* R&D efforts (Conductor Development Program, General Accelerator R&D GARD, university programs, etc.).

In this White Paper we propose the establishment of a similar ***Leading-Edge technology And Feasibility-directed Program (LEAF Program)*** to achieve readiness for a future collider decision on the timescale of the next decade.

Like for its predecessor, the LEAF Program would rely on, and be synergetic with, generic R&D efforts presently covered - in the US - by the Magnet Development Program (MDP), the Conductor Procurement and R&D (CPRD) Program and other activities in the Office of HEP supported by Early Career Awards (ECA) or Lab Directed R&D (LDRD) funds. Where possible, ties to synergetic efforts in other Offices of DOE or NSF are highlighted and suggested as wider Collaborative efforts on the National scale. International efforts are also mentioned as potential partners in the LEAF Program.

We envision the LEAF Program to concentrate on demonstrating the feasibility of magnets for muon colliders as well as next generation high energy hadron colliders, pursuing, where necessary and warranted by the nature of the application, the transition from R&D models to long models/prototypes. The LEAF Program will naturally drive accelerator-quality and experiment-interface design considerations. LEAF will also concentrate, where necessary, on cost reduction and/or industrialization steps.

The LEAF Program is foreseen to be a decade-long effort starting around ~2024-2025 to be concluded on the timescale of ~2034-2035. Based on the experience of the proponents, we suggest that the appropriate funding level for the LEAF Program should be ~25-30M$/year across the spectrum of participants (US National Laboratories & Universities).

## 2. INTRODUCTION

Following the successful start of the LHC in 2010 and the Nobel-prize discovery of the Higgs boson in 2012, the LHC has continued to help answer some of the key questions of the age such as the nature of dark matter; the existence of extra dimensions as well as continue to study the properties of the Higgs sector. An improvement to the LHC and its detectors, called HL-LHC [1], has been approved in 2016 to allow the full exploitation of the LHC in the third and fourth decades of this century and to allow unique research opportunities both in fundamental discoveries and in accelerator science.

The United States government is making an investment of more than $750M in the upgrade of the LHC to achieve the High Luminosities necessary to fully exploit the research frontier at the LHC energies. Part of these investments is supporting the construction of new Interaction Regions (IR) designed to increase tenfold the luminosity delivered to the detectors. The new machine, called the High-Luminosity LHC (HL-LHC), is presently in its construction phase. The US is contributing to the Accelerator part of HL-LHC through a DOE approved Project called the HL-LHC Accelerator Upgrade project (AUP), to be deployed at CERN for installation and commissioning of the HL-LHC in the 2024-2027 period [2].

It is an historical fact that the feasibility of the HL-LHC Project was made possible by the DOE investment in the LHC Accelerator R&D Program (LARP), mandated by OHEP and executed in 2003-2016 [3][4] as a directed R&D effort aiming at developing appropriate technologies for what was seen, in the early 2000's, as an inevitable upgrade of the LHC capabilities.

***In the same spirit as that under which LARP was initiated, this proposal of a Leading-Edge technology And Feasibility-directed Program is put forward with the aim of defining, exploring and demonstrating design and technologies to support an informed decision about the feasibility and start of Future Colliders endeavors in the next decade.***

## 3. GOALS

The circular nature of some of the Colliders under consideration for the future exploration of HEP frontiers naturally drives the focus to the study and development of advanced magnets in various configurations (dipoles & quadrupoles, solenoids, fast ramping magnets, etc.) although other technologies such as SRF or beam control/manipulation might become natural part of this Program as it evolves. The Circular Colliders considered for this feasibility-directed program include muon colliders [5][6] as well as high energy hadron colliders such as FCC-hh [7] or SppC. It is nevertheless expected that other machines such as planned electron-ion colliders will benefit from elements of this Program.

The Goal is to define and execute the appropriate steps from superconductor development, to design & development leading to the construction of model coils and magnets according to identified accelerator-quality needs for the muon collider or hadron collider, followed by the construction and characterization of real-size prototypes and focusing, on a case by case basis, on the most critical elements of the problem (from *Lab-based construction of one-of-a-kind, record-field solenoid or few Interaction Region (IR) quadrupoles* to *low-cost, run-of-the-mill, industrially produced Main Ring (MR) dipoles*).

Given the commonality of intents, the proponents believe that the goal is best achieved sharing resources among different funding agencies (DOE, NSF, etc.) and different Offices in the Funding agencies (OHEP, FES, ARDAP, …).

## 4. A FEW NUMBERS & TECHNICAL CHALLENGES

Superconducting Magnets (dipoles and quadrupoles) based on $Nb_3Sn$ technology have been demonstrated up to ~15T (single units) [25]. Hybrid solenoids using NbTi, $Nb_3Sn$ and HTS tape technology have been demonstrated up to 32T [14].

All the magnets mentioned above are produced in National Laboratories in single quantities or in "boutique" operations in quantities of few dozens in the 2020's, such as for the $Nb_3Sn$ focusing magnet for the HL-LHC.

A muon collider based on fast ramping magnet for muon acceleration [5][6] would require the magnets shown in Table 1. A very high energy hadron collider [7] would require the magnets shown in Table 2.

| Magnet Type | Field | Quantity |
|---|---|---|
| Production Target EHF Solenoid | ~40T | 1 |
| Cooling Channel VHF Solenoids | ~25T | Dozens |
| Cooling Channel HF Solenoids | ~4-15T | Hundreds |
| Fast Ramping Magnets | ΔB~2T and dB/dt=1000T/s | Few Hundreds |
| MR High Field Dipoles | ~8-12T | Few Hundreds |
| IR High Field Quadrupoles | ~15-16T | Dozens |

Table 1: Approximate count of magnets and field levels for a generic muon collider

| Magnet Type | Field | Quantity |
|---|---|---|
| MR High Field Dipoles | ~15-16T | Few Thousands |
| IR High Field Quadrupoles | ~15-16T or higher | Dozens |

Table 2: Approximate count of Magnets and field levels for a generic high energy hadron collider

The above considerations are exposing the two main *challenges* in addressing the feasibility of such future colliders in the next decade:

### 4.1. Industrialization Challenge

When needed quantities are in the "hundreds/thousands of units", industrialization is a must to maintain the necessary cost control and insure uniformity of deliverables. The cost control/reduction aspect was already identified as a challenge for magnet R&D in the 2014 P5 and associated HEPAP Accelerator R&D panel reports. This challenge applies to several beam-line magnets listed above (dipoles, fast-ramping magnets, cooling solenoids, etc) and the effort must involve Laboratories and Universities in the Design & Development (D&D) and prototyping phases, but needs to be followed by the demonstration of a feasible technology transfer and an appropriate industrialization process for the pre-series and series production phases

## 4.2. Field Level Challenge

When a high or very high magnetic field level is necessary to ensure the technical success of machine elements produced in small unit numbers (focusing IR magnets, production target solenoids, few dozens of Very High Field Solenoids, etc.) an approach based on Laboratories or Universities involvement from R&D to final production can be entertained given the inherent difficulties in technology transfer of high field magnets applications. In this respect, the present Whitepaper naturally includes the scope highlighted in the "Letter of Intent on Leading-Edge R&D effort finalized at the Interaction Regions of future Colliders" submitted in August 2020.

## 5. TEN YEAR PLAN

If approved and funding is started on the timescale of ~2024-2025, the first task for the LEAF Program is to develop a roadmap indicating the approach to a feasibility demonstration for muon colliders or high energy hadron colliders in the following decade.

Based on the experience of the proponents and assuming an inflation-adjusted LARP-like funding level (~25-30M$/year across the spectrum of involved participants) a plan based on the goals indicated below can represent a good starting point for discussion and further development.

| | 2024 | 2025 | 2026 | 2027 | 2028 | 2029 | 2030 | 2031 | 2032 | 2033 | 2034 |
|---|---|---|---|---|---|---|---|---|---|---|---|
| Extremely High Field Solenoid (>40T) | orange | orange | orange | orange | orange | orange | orange | yellow | yellow | blue | blue |
| Very High Field Solenoids (~25T) | orange | orange | orange | orange | yellow | yellow | yellow | yellow | yellow | blue | blue |
| High Field Solenoids (4-16T) | orange | orange | orange | yellow | yellow | yellow | yellow | green | green | green | green |
| | | | | | | | | | | | |
| Fast Ramping Magnets (~2T, 400 Hz) | orange | orange | yellow | yellow | yellow | yellow | yellow | green | green | green | green |
| | | | | | | | | | | | |
| High Field Dipoles (MR, ~12-16T) | orange | orange | orange | orange | orange | yellow | yellow | yellow | yellow | green | green |
| Very High Field Quadrupoles (IR, ~15-20T) | orange | orange | orange | orange | orange | orange | orange | orange | yellow | yellow | blue |

| | |
|---|---|
| orange | D&D and Small Scale Prototypes |
| yellow | Large Scale Prototpes |
| blue | National Lab Feasibility |
| green | Pre-series and Industrialization Feasibility |

Figure 1. Possible plan to define development phases of the magnets listed in Table 1 and 2.

## 5.1. D&D and Small Scales Prototype

The D&D effort would start from a definition of ranges for the necessary performance specifications for the colliders under consideration. The D&D phase would then continue to develop pilot runs for conductor and prototypes for magnets to downselect, at the appropriate maturity level, the spectrum of possible technical choices. ***The LEAF Program will focus on driving the D&D phases toward addressing definitions and specifications for accelerator-quality magnets, including - when appropriate - considerations of interfaces with the experiments.***

## 5.2. Large Scale Prototypes

The transition to full-scale prototype(s) is a necessary step to address all integration and performance elements that are typically not visible in small-scale elements. Also, the construction of full-scale prototypes enables an advanced development of tools and ergonomic solutions that would be critical in the production stages, irrespective of where the Production is executed. ***The LEAF Program major focus is expected to be on the Large Scale Prototypes.***

## 5.3. National Labs Feasibility

National Labs are well equipped to handle highly specialized production, from unique elements up to few dozens of units, as it is expected to be the case for Extremely High or Very High Field Solenoids or specialized IR magnets. In such cases, pursuing industrialization is not cost effective and ***the LEAF Program will concentrate on building the capabilities in the appropriate Laboratories or Universities.***

## 5.4. Pre-Series and Industrialization Feasibility

The need to prove the feasibility of industrialization is critical when the number of elements is in the hundreds to thousands. Early industrial involvement and – possibly – training of Industry representatives at National Labs is critical to ensure a favorable outcome. ***To achieve this goal, the LEAF Program will maintain constant awareness to the problematics of an industrialization process, starting from the magnet design and basic technology choices.***

# 6. ONGOING ACTIVITIES and RELATIONSHIP WITH the LEAF PROGRAM

## 6.1. HL-LHC Accelerator Upgrade Project (AUP)

The HL-LHC Accelerator Upgrade Project (AUP) is a 413.3b Project approved by DOE to support the US contribution to HL-LHC, namely 10 large-aperture IR low-β focusing $Nb_3Sn$ quadrupoles and 10 Radio Frequency Dipole (RFD) Crab Cavities, located in close proximity to the ATLAS and CMS experiments. The Project spans 2016 (CD-0) to 2028 (CD-4).

The whole HL-LHC upgrade, aiming at increasing by 10-fold the Luminosity delivered to the CMS and ATLAS experiments in the next decade (from 300 $fb^{-1}$ to 3000 $fb^{-1}$), was enabled by DOE investment in LARP, that coordinated the efforts from the US Labs involved in HEP to develop the technology needed for such an upgrade.

The AUP Project represents a "first" for humanity in its goal to build, commission and operate $Nb_3Sn$ magnets in a scientific accelerator such as the HL-LHC. AUP is now at approximately the 50% completion mark and will leave a legacy of invaluable components and tools (short length of superconductor, coil and magnet parts, assembly equipment, etc.) ***that can be devoted to the LEAF Program***.

Most importantly, the human capital (knowledge and resources) inherited from the execution of AUP and, especially, the training provided to younger scientists and engineers that have the privilege of participating in AUP activities, will be a core value on magnet expertise for this Nation and will be the main drivers, for several years in the future, of the LEAF Program. In return, ***the LEAF Program can build on this human capital by supporting continued talent development in US institutions by the establishment of Research Fellowships similar to the Toohig Fellowship supported by LARP in the last decade.***

## 6.2. DOE OHEP R&D (Magnets and Superconductor)

The DOE Office of High Energy Physics is presently supporting the MDP (Magnet Development Program) and the CPRD (Conductor Procurement and R&D) across several US National Labs. MDP has 4 main primary goals:

- Explore the performance limits of $Nb_3Sn$ accelerator magnets.

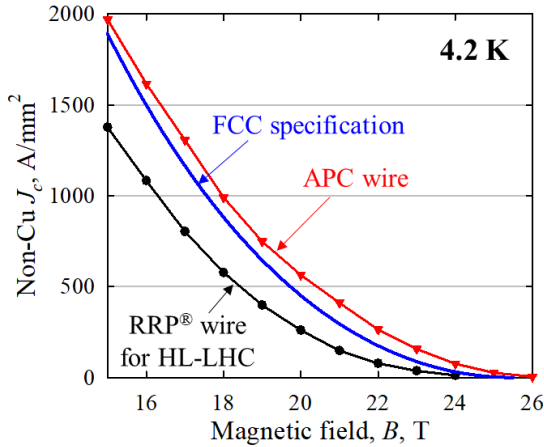

- Develop and demonstrate an HTS accelerator magnet with a self-field of 5T or greater (to be used with an $Nb_3Sn$ magnet in hybrid configuration)
- Investigate fundamental aspects of magnet design and technology
- Pursue $Nb_3Sn$ and HTS conductor R&D

The MDP program has enabled multiple ground-breaking advancements, such as development of several HTS research venues, demonstration of record-breaking accelerator-quality dipole to 14.5T [25], development of the CCT design and other technologies such as advance quench diagnostic (acoustics, fibers and quench antennas), exploring new insulation materials and epoxies, improving quench training, and recently identification of quench precursors using machine learning techniques. In addition, Early Career Awards and Lab Directed R&D funds have allowed the pursue of innovative efforts to improve the performance of $Nb_3Sn$ conductor using APC (Artificial Pinning Centers) to push the performance beyond what is currently commercially available for the HL-LHC AUP. Below we highlight some examples of these generic R&D efforts treated in more detailed in the MDP Whitepaper [8]. We also highlight how these MDP R&D efforts can naturally merge into the LEAF Program.

## 6.3. $Nb_3Sn$ Superconductor

Future energy-frontier hadron colliders (such as the FCC-hh) will require $Nb_3Sn$ conductors with performance much above the state of the art. Presently the rod-restack-process (RRP®) wires produced by Bruker OST set the benchmark for performance. Advancement, particularly in critical current density ($J_c$), are necessary because of their significant influences on the size and cost of the magnets and accelerators.

The $J_c$ of state-of-the-art $Nb_3Sn$ conductors has plateaued for two decades but is still significantly below performance considered to be necessary. Fermilab, in collaboration with Hyper Tech Research Inc., Ohio State University and Florida State University, is developing $Nb_3Sn$ conductors with new Nb-based alloys. The Fermilab approach aims to add artificial pinning centers (APC) based on an internal oxidation method [9][10] to the background pinning provided by $Nb_3Sn$ grain boundaries. Figure 2 shows the non-Cu $J_c$-B curves of a typical RRP®-type wire for the HL-LHC project and an APC wire, along with the FCC $J_c$-B specification. Apart from the higher $J_c$ at high field (above 10 T), the APC conductors also have an important extra benefit: due to the point pinning mechanism, they have much flatter $J_c$-B curves and thus lower $J_c$ at low field (e.g., below 5 T). This leads to significant reduction of magnetization at low field compared with non-APC $Nb_3Sn$. This is very desirable because the large magnetization of non-APC $Nb_3Sn$ at low field leads to large flux jumps, field errors, and

a.c. loss, which are all significant issues to solve for future high-field accelerator magnets [7]. The Florida State approach takes advantage of the raised recrystallization temperature in Nb-alloy, which facilitates reduction of the resultant $Nb_3Sn$ grains.

Figure 2. The non-Cu $J_c$-$B$ curves of a typical RRP®-type wire for the HL-LHC project and an APC wire, along with the FCC $J_c$-$B$ specification.

Development of $Nb_3Sn$ conductors with high specific heat, which increases their margin against quench and is promising to reduce the long training for $Nb_3Sn$ magnets [11] is also ongoing in the US. So far, it has been demonstrated that $Nb_3Sn$ wires with much higher specific heat can be produced without any issue, and the minimum quench energy can be significantly improved.

The technologies above are being actively considered n research supported by the LDRD and CPRD Programs. Industrialization of best techniques by Bruker-OST could occur over 2-3 years, with potential of make-to-specification billets thereafter at partial to full production scale. ***It is expected that long-length wires, with spool piece length above 1 km and cable unit length > 100 m can be available from industrial vendors for short magnet models or prototypes within the next 5 years, enabling LEAF in its development of high-performance $Nb_3Sn$ magnets for accelerators.***

## 6.4. HTS Superconductor

Presently, RE-$Ba_2Cu_3O_7$–δ coated conductors (REBCO) are among the most promising HTS materials available in 200-500 m lengths that have demonstrated engineering current densities of interest to high field magnets beyond $Nb_3Sn$. REBCO is a highly anisotropic superconductor that is only available in the tape form, which, however, does not require any heat treatment or post-processing [12]. In addition, REBCO tapes show remarkable mechanical properties in tension when the underlying material is Hastelloy [13]. In this respect, it is similar to NbTi and has a potential of becoming the workhorse HTS conductor for future magnet applications.

These properties made it nearly an ideal conductor for solenoid production, where HTS sections can be wound from a single tape as double-pancake coils with or without insulation, which allowed them to reach a world record 32 T field produced by an all-superconducting solenoid [14]. However, to generate the transverse (i.e. dipole and quadrupole) fields, the operation and protection requirements of future colliders place the optimum currents in the 10-20 kA range, which is orders of magnitude higher than what a single HTS tape can deliver. Moreover, test coils have revealed a range of conductor vulnerabilities to high strain, including propagation of slitting cracks and distortions due to screening currents [15][16] . REBCO conductor displays a wide range of property variations at 4K, partly because the primary quality control is carried out at 77 K and is not sensitive to some properties that do not scale with temperature [17].

This brings a need for the multi-tape HTS conductors. Due to the (flat) tape geometry, cabling techniques developed for Low Temperature Superconducting (LTS) materials cannot be directly applied to REBCO conductors. Currently, the most promising multi-tape REBCO conductor architectures are the Conductor on Round Core (CORC®) cables [18] developed by Advanced Conductor Technologies LLC and Symmetric Tape Round STAR wires [19] developed by AMPeers LLC.

Another promising HTS conductor is Bi-2212: a round, multifilamentary high temperature superconducting wire that can be made into a high current Rutherford cable. Recent years have seen significant development of high-temperature superconducting Bi-2212 wires and magnets in the US with record critical current density ($J_E$ of 1000 A/mm$^2$ at 4.2 K and 27 T) [20][21], record wire lengths (billets drawn in single 2 km pieces) and record performance in model magnets [22].

Between now and 2025, the U.S. Magnet Development Program centers on constructing and testing prototype dipole magnets (40 mm bore, 1 m long) with increasing magnetic field from 2T to 6.5T (standalone) and suitable to combine with large bore Nb$_3$Sn magnets (120mm, 11T, to be built by the US MDP) to explore hybrid magnet technology at 12-16 T. **The LEAF Program will direct the technology development toward large bore, high field 4-16 T muon capture solenoids, fast cycling magnets and very high field, 15-20 T, large bore, hybrid IR quadrupole magnets.**

## 6.5. Solenoid-like and Flat-Cable Magnets

The magnet technology based on flat cables with large aspect ratios was used in every superconducting collider built starting from Tevatron. It is ideal for NbTi and Nb$_3$Sn materials that can be readily formed into the keystoned Rutherford cables and wound as self-supporting coils of the Roman arch geometry. Use of the HTS conductors, especially in the round form, however, requires an additional support structure to hold the turns in the correct position as well as to reduce the transverse stresses on the conductors.

Canted-Cosine-Theta (CCT) magnet technology introduced in 2014 [23] offers the individual cable support, which is well suited for round conductors. Other approaches include the Stress Managed Cosine Theta (SMCT [24]) or the Conductor On Molded Barrel (COMB [40]). These technologies take on a different approach, which combines the individual conductor support inherent to CCT with the traditional cosine-theta coil geometry. An example of the COMB support structure is shown in Fig. 3 along with a mockup coil built to demonstrate the manufacturability. As can be seen from the picture, one solid structural part holds all the turns in two layers of the cable, which eliminates the disconnected pieces that must fit together with a high precision, as well as the layer-to-layer interface that can be a source of magnet training.

Presently the HTS magnet development based on CCT and COMB technologies is focused on short magnet models. This makes them well suited for studying the cable-specific effects, like bending degradation, current sharing, magnetization, quench propagation and protection, as well as scaling the transverse magnetic fields produced by HTS from the present target of 5 T into 10-15 T range and studying the hybrid operation of HTS/LTS magnets.

These activities are expected to last until 2030-2035 and will culminate with several demonstration models, both standalone and as parts of hybrid, REBCO/Nb$_3$Sn systems. The scale-up of the HTS magnet length is synergistic to and can be done in parallel with the production of the long Nb$_3$Sn magnet prototypes in the 2030-2035 time frame. *This goal will need a dedicated effort such as the LEAF Program as the magnet length scaling requires resources that go well beyond the MDP level of funding.*

Figure 3. A COMB dipole magnet consisting of two half-coils (center) and the mock-up coil (top and bottom).

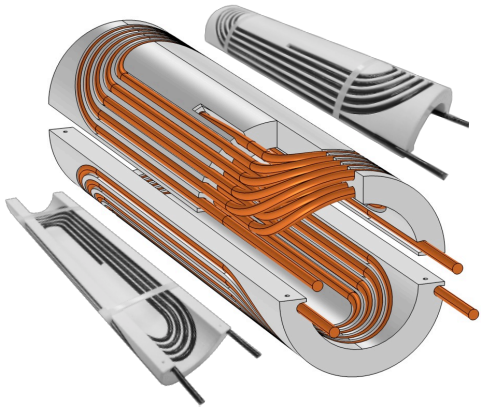

## 6.6. High Field MR Dipoles and IR Quadrupoles

High Field Main Ring dipoles and Interaction Region Quadrupoles are a must for any of the machines considered in this whitepaper. Prototype of an accelerator-quality Nb$_3$Sn dipole has reached ~14.5T [25] while small-quantities production (few dozens) of accelerator-quality quadrupoles (HL-LHC) are presently pushing the field limit to the ~12T region [2].

However, the present cost of Nb$_3$Sn accelerator magnet needs to be significantly reduced in other to allow production of these magnets in the hundreds or thousands as needed by the colliders mentioned here. Robust design is critical as well since automation and industrialization are going to be key factors in this process. Therefore, it's critical to develop first a technology compatible with automation and industrialization of Nb$_3$Sn accelerator magnets.

A "next generation" technology must be developed in collaboration between national laboratories (who master the 1$^{st}$ generation technology used in LARP and HL-LHC) and industry (who master automation and industrialization). Significant cost reduction (factor of 2 or higher) is possible because the 1$^{st}$ generation technology was optimized for achieving performance while breaking new ground. In the second half of this decade, the experience and know-how being built in national laboratories through the production of the Nb$_3$Sn IR quadrupoles for HL-LHC are going to allow a very fruitful collaboration with industry. A phased program for the development of this "next generation" technology for the fabrication of cost-effective Nb$_3$Sn accelerator dipoles and quadrupoles in the 12-14 T range has been presented in this Snowmass process [26].

Focusing quadrupoles benefit from the lower number of magnets needed in a Collider application. Hybrid solutions (LTS/HTS) can be developed to benefit from the "next generation" technology described above for the LTS section in combination with a HTS solution made possible in advancement covered in the HTS Superconductor Section.

Finally, when applications for muon collider are considered, the problem of heat deposition from muon decays in the superconducting coils of the magnets has been studied in the past [27] and can be addressed by sophisticated protection systems in the MR and IR to bring the peak power density in the coils well below the quench limit and reduce the dynamic heat deposition in the cold mass of the magnets by a factor of 100. Solutions based on protection system consisting of tight tungsten masks in the magnet interconnect regions and elliptical tungsten liners in the magnet aperture optimized for each magnet can be employed. ***These studies will have to be tailored for the final design on the Collider under consideration and are a natural element of the LEAF Program.***

## 6.7. Fast Ramping Magnets

Next generation HEP facilities such as muon colliders [5][6], future circular colliders [7][28] and high-intensity proton synchrotrons for neutrino research [29][30][31] accelerators demand substantially faster cycles of beam acceleration than available at present. To-date, the highest ramping rates achieved in the operational superconducting accelerator magnets based on the LTS (NbTi) are about 4 T/s [32][33], a limitation caused by a very narrow allowable operational temperature margin.

Fast-ramping HTS-based magnets offer a cost-effective solution for many future particle accelerators mentioned above but especially for the acceleration of the short-lived particles such as muons. The AC losses in the fast-ramping accelerator magnet are due to power losses in both the magnet energizing conductor and the magnetic core. The power losses in the magnetic core can be reduced by using thin laminations. The power losses in the conductor can be reduced by minimizing both its mass and exposure area to the ramping magnetic field descending from the core. Application of an HTS superconductor allows to strongly minimize magnet cable mass and size, and as a result also the size and mass of the magnetic core. Very importantly, however, the HTS conductor can be set to operate at 5 K, well below its critical temperature of ~30 K, providing in this way a wide operational temperature margin and facilitating the temperature-based quench detection and protection systems. The conceptual design of an HTS-based accelerator magnet is shown in Fig. 1.

Fig. 4. Conceptual design of a dual-bore HTS-based accelerator magnet.

The vertically arranged 2-bore magnet is powered with a single conductor coil. The magnetic fields in the beam gaps are of opposite orientation allowing to accelerate oppositely charged particles ($\mu^+\mu^-$, $e^+e^-$) in the same direction using the unipolar magnetic field waveform. For acceleration of the same charge particles ($\mu^+\mu^+$, $e^-e^-$, pp) a bipolar magnetic field waveform can be used.

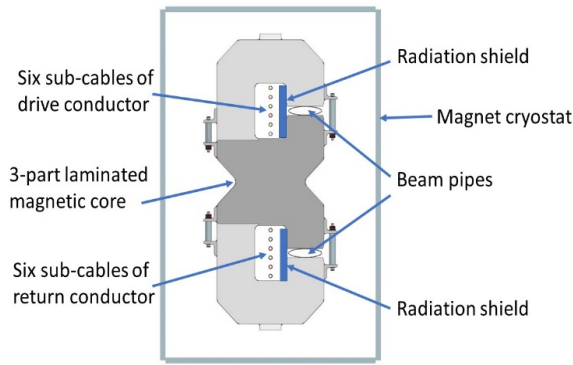

A prototype HTS-based accelerator magnet of 0.5 m length and two beam gaps of 100 mm (horizontal) x 10 mm (vertical) was successfully tested [34]. With up to 0.5 T magnetic field in the beam gaps the maximum dB/dt rates for unipolar and bipolar current waveforms were 274 T/s and 289 T/s, respectively. Within the temperature error of +/- 0.003 K no rise of the conductor temperature was observed for the magnet excitation current of 3 kA at current pulse repetition rates of 1 Hz to 14 Hz leading to the upper limit on the cryogenic power loss of less than 0.1 W. Preparations are now underway to increase this test magnet magnetic field to 0.9 T and the ramping rates up to (500-600) T/s.

Future goals in the next 2 years include an upgrade of the present HTS test magnet to 2T field and the dB/dt rates up to 800 T/s and are described in a Whitepaper to this Snowmass process [35]. In the longer term (3-6 years), goals should include the design, *construction and power test a long prototype magnet as required for the muon accelerator and the initiation of a possible industrialization process, under the LEAF Program Umbrella*

## 6.8. Facilities and Technical Support

All the National Laboratories involved in the DOE HEP efforts for leading edge magnet technology bring to the effort world class facilities supported by the Agency investments in the last ~40 years since the technology has been introduced and proven in the High Energy Physics field itself.

For example, LBNL, BNL, and FNAL have world class magnet and cable construction and test facilities that are currently being used for magnet testing of the AUP Project and for R&D Magnets and Conductor testing.

LBNL and FNAL are equipped with tools for Rutherford cable fabrication. LBNL, for example, has a 60-strand cabling machine with a capacity of 200 kg. The machine includes a world-class suite of diagnostics and cabling QC tools such as an in-line cable parameters measurement engine (CME), an imaging system that can capture all four sides of every inch of the manufactured cable through the entire run, a cryocooler system for measuring extracted strand residual resistance ratio. The machine is presently used for $Nb_3Sn$ cables for AUP, and $Nb_3Sn$/Bi2212 cables for magnet R&D and international collaborators.

BNL developed an HTS cable and hybrid magnet testing in the 10T common coil low-cost, rapid-turn-around technology test facility, that can be used for the LEAF Program as well. The 10 T common coil dipole test facility is currently providing a field of 0 -10 T at ~4 K and up to 10 kA in cable if connected in series with the magnet and upgrades to the facility are in the plan.

As an example of Magnet test facilities, BNL currently has 3 vertical test dewars and 2 horizontal test stands available for testing magnets. The vertical test facilities can operate at currents ranging from 8.5 to 30kA, with depths ranging up to 6.1 meters deep, and temperatures ranging from 1.9-4.5 K. . These test facilities will be used for testing the IR region magnets for the Electron Ion Collider but can also be used for the LEAF Program as well.

Similar facilities exist at FNAL, including a horizontal test facility operating at 1.9-4.5K, to be used for the AUP Project but available upon its completion in ∼2026, and 3 vertical test stand facilities, one of them (HFVMTF) is in construction and it is built in collaboration with the Fusion community to house a high field magnet (∼15T) to provide a background field for HTS cable studies supporting both HEP and FES. Expected to be ready by the end of 2023, it will be capable to test hybrid magnets with different configuration of powering.

*All national Labs involved in AUP will also bring to the LEAF Program extensive mechanical knowledge and assembly tooling to apply to the LEAF Program magnet construction efforts.*

# 7. SYNERGIES, COLLABORATIONS AND INDUSTRIALIZATION

## 7.1. Other DOE Offices and NSF

The efforts on High Field Solenoids supported by NSF are a ***clear opportunity of collaboration and synergy with the LEAF Program***. Past efforts using Bi-2212 and REBCO have pushed to 35 T and 45.5 T respectively as test coils in a background field provided by a resistive magnet.

An ongoing effort aiming at a 40 T solenoid [41] will produce a final design, including construction drawings, and an implementation plan. This effort will be followed by an NSF future proposal for construction of the 40 T superconducting magnet in the next decade. There is high potential for synergy and collaboration to address the needs of a future muon collider within the LEAF Program.

## 7.2. International Collaboration

High Energy Physics, and the major technologies associated with a future energy-frontier facility, are international endeavors and it is critical that progress and efforts are communicated and shared in the interest of the community. Accelerator magnet R&D has been pursued at laboratories around the world for decades; however the convergence of interests towards candidate facilities such as the FCC and the Muon Collider whose physics reach is directly linked to feasible magnet designs has already led to coordination of activities, first in the US with the formation by DOE-OHEP of the US Magnet Development Program (MDP) in 2016, and most recently by Europe with the formation of the High Field Magnet (HFM) program [36].

As development efforts become more focused on feasibility and industrialization, as is the focus of this whitepaper proposal, the potential value for coordination of efforts - beyond communication and sharing of data as is done in the existing R&D programs - grows. DOE has mechanisms in place to enable such collaborations with CERN, for example, and we would expect the scope and mechanisms for coordinating activities would be formalized by a Memorandum of Understanding. We note that the scope of the European HFM program includes significant aspects similar to the scope of the US MDP, but also includes scope that is well aligned with this proposal, that are currently beyond the purview of MDP.

Similarly, the US magnet programs have long collaborated with KEK-Japan, and scientists from KEK have significant experience in working with Japanese industry in magnet technologies; that experience is highly relevant to the scope of this proposal, and we anticipate a robust collaboration with KEK within this program.

These collaborative efforts on R&D have matured from and built on a multi-year international collaboration of accelerator magnets for international facilities. The US contribution to HL-LHC has been described before [2]. An example of a US contribution to a Japanese facility, BNL has developed and built the current set of IR corrector magnets currently in operation at SuperKEKB [37]. These correctors were developed using the unique BNL direct wind technology and all the magnets reached operating current with no training. BNL has ongoing collaborations with KEK under the US-Japan collaboration developing unique vibration measurement techniques and also novel, compact superconducting spin rotator units, that are basically drop-in replacements for four SuperKEKB ring dipoles.

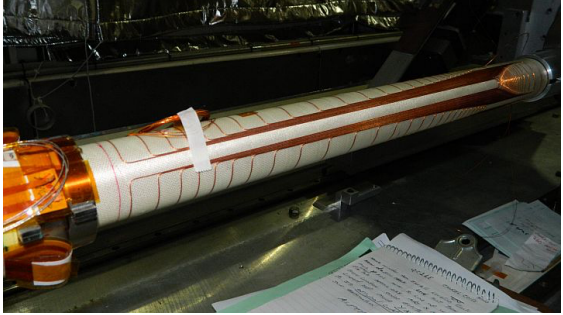

Figure 5: SuperKEKB IR Dipole Corrector

### 7.3. Industry

Industry involvement in the process of superconducting wire or tape fabrication is by now a well-established business in support of magnets construction. Even for generic R&D efforts such as those entertained by MDP, reliance on industrial superconductor vendors is a must.
*(One sore note on the conductor front is the discontinuation of Al-clad cable production by several companied around the world. While this is mostly a problem for detector magnets, future availability of aluminum-stabilized conductors can only be guaranteed by investments in infrastructure at one of the national institutions).* For production of accelerator magnets superconductor, several companies are lining up as potential industrial partners. In no particular order they include: Bruker OST (advanced $Nb_3Sn$, Bi-2212), Bruker EAS (advanced $Nb_3Sn$), JASTECH (advanced $Nb_3Sn$), Furukawa ($Nb_3Sn$ and Al-clad conductors), Hitachi ($Nb_3Sn$ and Al-clad conductors), Western Superconductor Technologies (advanced $Nb_3Sn$), qualified REBCO suppliers for fusion (SuperPower-Furukawa, Sunam, SuperOx, Fujikura, Shanghai ST, THEVA) and Sumitomo (Bi-2223 conductor).

The situation is different for the industrial production of magnets. While, ultimately, the industrial production of large quantities of magnets will be the responsibility of the entity executing the eventual collider, a Feasibility demonstration like the LEAF Program will need to steer the design and prototyping of any component slated to be produced in industry toward a solution that can be executed.

Several efforts in the past have involved industrial participation in the production of magnets for HEP applications. Some of the involvements have been extremely proficient, while other have created a treasure trove of Lessons Learned.

For very large productions, a coordinated effort must be made within the program to transfer technology to industrial partners. Early industry involvement, training at National Labs, a constructive dialogue and cross-pollination of ideas, especially during the early design phase, are all key elements to a successful outcome of the program. Presence of National Labs experts at the industry partner is also crucial to ensure quality and timely deliveries. Below, we consider the paths adopted by two successful Project in the past: magnet for RHIC and Transport Solenoid for the Mu2e Experiment.

A strategy of early involvement and "build to print" was adopted by BNL for the RHIC magnet production, which is a perfect study-case [38][39] in view of the "Industrialization Challenge" mentioned above for the LEAF Program.

BNL built and tested several (~a dozen) full length prototype dipole magnets to ensure manufacturability and satisfaction of performance requirements.

The contract was structured in a manner to encourage vendor participation by placing early risk with BNL. As many interested parties had little or no superconducting magnet construction experience, the first phase of the contract, which included all tooling design and construction as well as the fabrication of the first thirty magnets, was a "cost plus fixed fee" structure. In this way, cost issues would not result in pressure to compromise the quality of tooling or magnets. This proved crucial to the success of the contract, as differences in culture resulted in the selected vendor initially discounting the precise tolerances specified for tooling and magnets, thereby initially underestimating the cost of tooling and magnet components.

The remaining phases of the contract were structured as "fixed price incentive fee", again as an encouraging motivational tool for the vendor. In these phases, any cost savings or cost increases to the contract were shared between BNL and the vendor at values of 25% and 75%, respectively.

In addition, BNL maintained responsibility for parts supply (superconducting cable, stainless steel beam tubes, quench protection cold diodes and yoke steel, etc.) and organized Industrialization Open House to share experience and know-how.

Once the vendor was selected work began immediately on tooling designs and magnet materials procurements. Construction of prototypes at the vendor started in earnest, performed by the vendor's technical staff and witnessed by both the vendor's engineering staff and by BNL technical staff. After work procedures were approved, and tooling constructed, BNL technical and engineering staff visited the vendor as needed to assist with the debugging and commissioning of the tooling and equipment, and subsequent use for the first production magnet. The fulltime presence of a BNL representative at the vendor to facilitate implementing minor change requests and monitor progress was essential during production.

A similar approach was used by another successful magnet construction effort at a vendor considered here, namely the Transport Solenoid for the Mu2e experiment at FNAL. This effort used the same approach of BNL for components, with the Lab managing and overseeing progress at the various industrial partners while owning some of the coordination and final component assembly work. For the Mu2e Transport Solenoid, all superconducting strands and cables, cold mass elements, thermal shields and cryostats were manufactured by separate domestic and international industrial partners, while cold tests and final assembly is carried out at a National Laboratory.

***The LEAF Program will improve the probability of success for the final industrialization by relying on approaches for final D&D and prototyping similar to those mentioned above.***

## 8. CONCLUSION

The introduction of superconducting technology on large scales following the construction of the Tevatron in the ~70's has been a watershed moment for humanity, supported by world-leading infrastructure and capabilities in the US.

The US leadership on the front of superconducting magnet technologies has been preserved in the recent past, through the support of generic and direct magnet R&D programs which have enabled, among other things, the application of $Nb_3Sn$ technology to the HL-LHC research program in the framework of international partnership.

The ***Magnet LEAF Program*** proposed in this Whitepaper is a necessary step to maintain US leadership, infrastructure and human talent on the front of advanced technologies for future research facilities – irrespective of their geographical location.